# Push- and Pull-based Epidemic Spreading in Networks: Thresholds and Deeper Insights

Shouhuai Xu, University of Texas at San Antonio
Wenlian Lu, Fudan University
Li Xu, University of Texas at San Antonio

Understanding the dynamics of computer virus (malware, worm) in cyberspace is an important problem that has attracted a fair amount of attention. Early investigations for this purpose adapted biological epidemic models, and thus inherited the so-called homogeneity assumption that each node is equally connected to others. Later studies relaxed this often-unrealistic homogeneity assumption, but still focused on certain power-law networks. Recently, researchers investigated epidemic models in *arbitrary* networks (i.e., no restrictions on network topology). However, all these models only capture *push-based* infection, namely that an infectious node always actively attempts to infect its neighboring nodes. Very recently, the concept of *pull-based* infection was introduced but was not treated rigorously. Along this line of research, the present paper investigates push- and pull-based epidemic spreading dynamics in arbitrary networks, using a Non-linear Dynamical Systems approach. The paper advances the state of the art as follows: (1) It presents a more general and powerful sufficient condition (also known as epidemic threshold in the literature) under which the spreading will become stable. (2) It gives both upper and lower bounds on the global mean infection rate, regardless of the stability of the spreading. (3) It offers insights into, among other things, the estimation of the global mean infection rate through localized monitoring of a small *constant* number of nodes, *without* knowing the values of the parameters.

## 1. INTRODUCTION

Kephart and White [Kephart and White 1991; 1993] initiated the study of computer virus dynamics by adapting certain homogeneous biological epidemic models [McKendrick 1926; Kermack and McKendrick 1927; Bailey 1975; Anderson and May 1991; Hethcote 2000] where each individual has equal contact to the others. Recently, researchers (mainly statistical physicists) weakened the homogeneity assumption by considering spreading in heterogeneous networks, but mainly focused on spreading in power-law networks (cf. [Moreno et al. 2002; Pastor-Satorras and Vespignani 2001; 2002; Newman 2003; Barrat et al. 2008]). Very recently, computer scientists investigated the modeling of spreading dynamics in *arbitrary* networks (i.e., not just power-law ones) [Wang et al. 2003; Ganesh et al. 2005; Chakrabarti et al. 2008]. These models considered *push-based* spreading (attacks or infection), by which a virus always actively attempts to infect the susceptible computers. Push-based spreading (and thus these models) cannot accommodate *pull-based* attacks such as "drive-by download" [Provos et al. 2007] that a susceptible computer can get infected by connecting to a compromised website. This inspired researchers to introduce the more realistic *push-* and *pull-based* models in [Li et al. 2007], which however conducted mainly a simulation-based study and did not offer deep insights. This calls for a systematic and analytic study of push- and pull-based epidemic spreading dynamics in arbitrary networks, which motivates the present paper.

Although it is intuitive that push- and pull-based spreading dynamics models can accommodate attacks that are not captured by push-based models, one may still ask whether this is actually the case. This is a legitimate question because the abstraction of push-based spreading may be able to accommodate pull-based spreading as well (e.g., by attempting to adjust some parameters in push-based models). However, it turns out not to be the case because pull-based spreading and push-based spreading are two mechanisms that exhibit fundamentally/conceptually different behaviors. Specifically, pull-based spreading can accommodate the outside environment of a network in question (e.g., a compromised website does not belong to the network system in which the push-based epidemic spreading takes place); whereas, push-based spreading cannot accommodate the outside environment. It therefore appears to be necessary to introduce a new type of parameter, called $\alpha$, which abstracts the probability that a node gets infected because of its own reasons such as accessing a malicious website. As such, push-based spreading dynamics corresponds to the case of $\alpha = 0$, and push- and pull-based spreading dynamics corresponds to the case of $\alpha > 0$. In particular, when compared with the case of $\alpha = 0$, in the case of $\alpha > 0$ the spreading would never die out and the dynamics is much more difficult to analyze.

### 1.1. Our Contributions

First, we present a general sufficient condition (also known as epidemic threshold in the literature) under which the push- and pull-based epidemic spreading will become stable. This result supersedes its push-based counterpart given by [Chakrabarti et al. 2008] (ACM TISSEC'08).[1] In other words, our sufficient condition is more general and powerful. We also give a more succinct variant sufficient condition which, among other things, allows to tune (in a quantitative rather than qualitative fashion) the model parameters so as to assure that the spreading will become stable. This has important implications. For example, we can first force the spreading dynamics to become stable and then use the global mean infection rate (i.e., the rate or portion of infected nodes) as an index to indicate the overall security of a network and to support higher-level decision-making.

Second, we notice that it may not be always possible to force the spreading to become stable (e.g, the cost may be prohibitive). When the spreading is not stable (which does not necessarily mean there is an out-break in the setting of this paper with $\alpha > 0$), it would be important to know what are the worst-case and best-case scenarios in terms of the global mean infection rate. Towards this end, we give upper and lower bounds of the global mean infection rate. The bounds are actually applicable no matter the spreading dynamics is stable or not, and are heuristically tight in some cases. Unfortunately, we are currently unable to precisely pin down the threshold above which the upper bound would be tight, and below which the lower bound would be tight. Still, we note that results related to non-stable scenarios are very difficult to obtain in general.

Third, we offer the following deeper insights that are, to our knowledge, the first of their kind when considering arbitrary networks. When the spreading is stable,

— we show how node degree governs node infection rate;
— we give a condition under which the popular *mean field* approach is applicable to study push- and pull-based dynamics in arbitrary networks;

---

[1]As a side-product, we point out an error in [Chakrabarti et al. 2008], by which their sufficient condition was mistakenly claimed to be "necessary" as well. The details will be given in Section 4.1.

— we show how the global infection rate can be estimated by monitoring a *small* number of nodes selected in a certain fashion *without* knowing the values of the parameters and *independent of* the size of the network.

The rest of the paper is organized as follows. In Section 2 we discuss the related prior works. Because our work is built on top of the concept/model of push- and pull-based spreading dynamics introduced in [Li et al. 2007], we briefly review it in Section 3. In Section 4 we present some sufficient conditions (i.e., epidemic thresholds) under which the spreading will become stable, and present some upper and lower bounds on the global mean infection rate. In Section 5 we present some deeper insights that are especially useful in the absence of certain system information. We conclude the paper in Section 6 with open questions for future research.

## 2. RELATED WORK

The study of epidemic spreading dynamics on complex networks has become a hot topic. From a Computer Science perspective, this perhaps has been catalyzed by the ubiquitous presence of real-life networks, including social networks, Internet, and email networks. The study has been inherently multidisciplinary, and researchers have investigated approaches such as Epidemiology, Dynamical Systems, Stochastic Processes, Percolation Theory, Generating Functions, Algebraic Graph Theory, and Statistical Physics. As such, a thorough and complete review of the literature is clearly beyond the scope of the present paper. Instead, we here focus on a specific perspective, and classify the most-relevant existing studies of computer virus/malware dynamics into the following three generations. (Because our focus is not about immunization, we do not aim to discuss the vast volume of literature in this subfield as well.)

The first generation, initiated by Kephart and White [Kephart and White 1991; 1993] and followed by numerous variant studies such as [Zou et al. 2002; Zou et al. 2003], focused on homogeneous networks. Such homogeneous models essentially require the underlying communications to form complete graphs. This is unrealistic because some computers (or IP addresses) are blocked from accessing certain other computers (or IP addresses), and that malware spreading does not necessarily exploit the fully random scanning strategy [Chen and Ji 2005]. Moreover, the spreading process that exploits social networks (including email communication networks) are certainly not complete graphs, but rather follows the social networks' topologies.

The second generation focused on specific heterogeneous (i.e., power-law) networks, for which we refer to the most-recent excellent books [Barrat et al. 2008; Newman 2010] for a thorough treatment on the topic and for the large body of literature. From a technical perspective, we here highlight that there are mainly two approaches to tackling such heterogeneity. One approach is to exploit the specific properties of the degree distributions, especially the Barabasi-Albert power-law degree distribution and the affiliated preferential-attachment model [Barabasi and Albert 1999], which however are not without limitations (see, for example, [Faloutsos et al. 1999; Willinger et al. 2009]). The large volume of research results in this widely known approach were nicely summarized in the 2008 book of Barrat et al. [Barrat et al. 2008]. The other (perhaps less known) approach is to exploit generating functions [Wilf 1994]. This approach has successfully made a connection to the Percolation Theory [Callaway et al. 2000; Newman et al. 2001; Newman 2007] and has successfully identified epidemic thresholds. Moreover, this approach is applicable to random graphs for a given degree sequence (which can correspond to power-law distributions) as introduced by Molloy and Reed [Molloy and Reed 1995; 1998]. A systematic treatment of this approach can be found in the 2010 book of Newman [Newman 2010].

The third generation considers arbitrary heterogeneous networks. The earlier investigations focused on push-based spreading [Wang et al. 2003; Ganesh et al. 2005; Chakrabarti et al. 2008]. These investigations made a clear connection between the fate of push-based dynamical processes in arbitrary networks and the algebraic properties of graphs (more specifically, the largest eigenvalue of the adjacency matrix). The present paper enriches the dynamical processes by considering both push- and pull-based epidemic spreading — a concept first introduced in [Li et al. 2007], which however did not go deep enough analytically. To be more precise, the two papers that are the predecessors to the present work are: [Chakrabarti et al. 2008], which presents the state-of-the-art epidemic threshold in push-based model, and [Li et al. 2007], which introduced the push- and pull-based model but without giving deep results. We will demonstrate how our results supersede theirs.

Finally, we note that the epidemic spreading dynamics model considered in the present paper is essentially the so-called Susceptible-Infection-Susceptible (SIS) dynamics in nature [Bailey 1975], which are very different from gossip processes (e.g., [Demers et al. 1987; Karp et al. 2000; Kempe et al. 2001; Kempe and Kleinberg 2002; Shah 2009]), which are essentially Susceptible-Infection (SI) dynamics in nature.

### 3. A BRIEF REVIEW OF THE PUSH- AND PULL-BASED MODEL

We consider push- and pull-based epidemic spreading in complex networks, which are naturally represented as graphs. Because any topology could be possible and we want to draw insights that are widely applicable, we do not make any assumption on the networks' topologies, according to which the epidemic spreading takes place. In the rest of this section we review the push- and pull-based model introduced in [Li et al. 2007].

Specifically, we consider a finite network graph $G = (V, E)$, where $V = \{1, 2, \ldots, n\}$ is the set of nodes or vertices and $E$ is the set of edges or arcs. Denote by $A = [a_{vu}]$ the adjacency matrix of $G$, where $a_{vv} = 0$ and, if and only if $(u, v) \in E$ for $v \neq u$, we have $a_{vu} = 1$. The graph $G$ captures the topology according to which the spreading takes place, where $u$ can directly infect $v$ only when $(u, v) \in E$. For undirected graph, let $\deg(v)$ denote the degree of node $v$ in $G$. The discussion in the paper focuses on undirected graphs, but can be adapted to the setting of directed graphs by replacing $\deg(v)$ as the in-degree of $v$ in directed graph $G$.

Consider a discrete time model with time $t = 0, 1, 2, \ldots$. At any time $t$, a node $v \in V$ is either **susceptible** or **infectious**. (In this paper, "infectious" and "infected" are used interchangeably.) Denote by $s_v(t)$ the probability that $v \in V$ is **susceptible** at time $t$, and $i_v(t)$ the probability that $v \in V$ is **infectious** at time $t$. The model invariant is $s_v(t) + i_v(t) = 1$ for any $t \geq 0$. Moreover, a **susceptible** node $v$ may become **infectious** at a single time step because of push- or pull-based infection, and an **infectious** node may become **susceptible** because of defense or cure. This explains why our model is SIS (Susceptible-Infection-Susceptible) in nature, except that it considers both push- and pull-based infections. To accommodate push- and pull-based spreading, we use the following parameters:

— $\alpha$: The pull-based infection capability, namely the probability a **susceptible** node becomes **infectious** at a discrete time step because of its own activity (e.g., connecting to a malicious website which may not belong to $G$).
— $\gamma$: The push-based infection capability, namely the probability that an **infectious** node $u$ successfully infects a **susceptible** node $v$, where $(u, v) \in E$. In the case of undirected graph, it is natural that $\gamma_{uv} = \gamma_{vu}$ for all $(u, v) \in E$.
— $\beta$: The cure capability, namely the probability that an **infectious** node becomes **susceptible** at a single time step.

The main notations used throughout the paper are summarized below.

| | |
|---|---|
| $\lambda_{1,A}, \ldots, \lambda_{n,A}$ | the eigenvalues of the adjacency matrix A of network graph $G$ with $\lambda_{1,A} \geq \ldots \geq \lambda_{n,A}$ (in modulus) |
| $\lambda_{\max}(M)$ | the largest eigenvalue (in modulus) of matrix $M$ |
| $M^\top$ | the transpose of matrix (or vector) $M$ |
| $s_v(t)$ | the probability that node $v$ is **susceptible** at time $t$ |
| $i_v(t)$ | the probability that node $v$ is **infectious** at time $t$ |
| $\bar{i}(t)$ | the global mean infection rate (or probability) $\frac{1}{|V|}\sum_{v \in V} i_v(t)$; $\bar{i} = \lim_{t \to \infty} \bar{i}(t)$ is also used in the case the spreading is stable |
| $\bar{i}[t_0, t_1]$ | the average of $\bar{i}(t)$ over time interval $t \in [t_0, t_1]$ |
| $\langle r \rangle$ | the mean of random variable $r$ |
| $\alpha$ | the probability that a node becomes **infectious** at a single time step because of pull-based infection |
| $\beta$ | the probability that an **infectious** node becomes **susceptible** at a single time step |
| $\gamma$ | the probability that a **susceptible** node $v$ is infected by an **infectious** neighbor $u$ where $(u, v) \in E$ at a single time step |
| $\delta_v(t)$ | the probability that a **susceptible** node $v$ becomes **infectious** at time $t + 1$ because of its **infectious** neighbors $\{u : (u, v) \in E\}$ at time $t$ |
| $I_n$ | the $n \times n$ identity matrix (i.e., $a_{jj'} = 1$ if $j = j'$, and $a_{jj'} = 0$ otherwise) |

Figure 1 depicts the state transition diagram [Li et al. 2007], according to which node $v \in V$ changes its state. As in [Chakrabarti et al. 2008], we may assume that the (neighboring) nodes' states are independent. Then, at time $t+1$, $v$ may become **infectious** because of pull-based infection with probability $\alpha$, or because of push-based infection with probability $\delta_v(t)$, where

$$\delta_v(t) = 1 - \prod_{(u,v) \in E} [1 - \gamma i_u(t)]. \qquad (3.1)$$

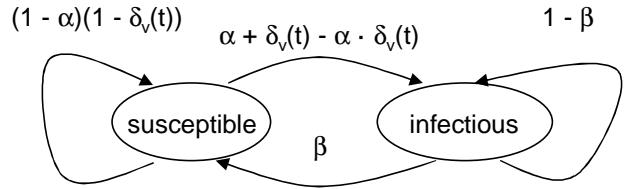

Fig. 1. The state transition diagram for node $v \in V$

As a result, the master equation of the nonlinear dynamical system is [Li et al. 2007]

$$\begin{cases} s_v(t+1) = [(1-\alpha)(1-\delta_v(t))]\, s_v(t) + \beta i_v(t) \\ i_v(t+1) = [1 - (1-\alpha)(1-\delta_v(t))]\, s_v(t) + (1-\beta)i_v(t), \end{cases}$$

namely

$$i_v(t+1) = \left[1 - (1-\alpha)\prod_{(u,v) \in E}(1 - \gamma i_u(t))\right](1 - i_v(t)) + (1-\beta)i_v(t). \qquad (3.2)$$

Note that the above master equation preserves the invariant $s_v(t) + i_v(t) = 1$.

## 4. EPIDEMIC THRESHOLDS AND BOUNDS ON INFECTION RATE

In this section we address the following questions in the above push- and pull-based model.

— What are the sufficient conditions (i.e., epidemic thresholds) under which the spreading will become stable *without* resorting to any approximation (which is often used for analyzing nonlinear dynamical systems)? We address this question in Sections 4.1-4.2.
— Can we bound the node infection rate *without* requiring the spreading to become stable (Section 4.3)? Such results would be useful especially when it is impossible or prohibitive to "manipulate" or force the spreading dynamics to become stable.

In addition, we will discuss the applications of these theoretic results (Section 4.4).

### 4.1. A General Epidemic Threshold

Unlike the push-based model, where $\alpha = 0$ and thus $\lim_{t \to \infty} i_v(t) = 0$ for all $v$ is a trivial equilibrium state, we do not have such a leverage for push- and pull-based model because $\alpha > 0$ and thus it is almost always true that $i_v(t) > 0$, meaning that the spreading would not die out. Although we are unable to point out where the equilibrium state is, we show that there must be an equilibrium state (i.e., existence) that is exponentially globally stable.

THEOREM 4.1. (a general sufficient condition under which the spreading will become stable) *Let $[i_1^*, \cdots, i_n^*]$ be an equilibrium of nonlinear dynamical system (3.2), $H = diag[h_v]_{v=1}^n$ with*

$$h_v = \left| -\beta + (1-\alpha) \prod_{(u,v) \in E} (1 - \gamma i_u^*) \right|$$

*and the other parameters be specified as above. If*

$$\lambda_{\max}(H + \gamma(1-\alpha)A) < 1, \tag{4.1}$$

*then system (3.2) is globally exponentially asymptotically stable, namely that $\lim_{t \to \infty} i_v(t) = i_v^*$ holds for $v = 1, \cdots, n$ regardless of the number of the initially infected nodes.*

PROOF. Let $I(t) = [i_1(t), \ldots, i_n(t)]^\top$. Then Eq. (3.2) indicates $I(t + 1) = f(I(t))$ for some continuous function $f : [0, 1]^n \to [0, 1]^n$. Since $[0, 1]^n$ is convex, Brouwer's fixed point theorem [Yoshida 1971] says that there exists at least one equilibrium $i^* = [i_1^*, \cdots, i_n^*]^\top$, i.e.,

$$i_v^* = \left[ 1 - (1-\alpha) \prod_{(u,v) \in E} (1 - \gamma i_u^*) \right] (1 - i_v^*) + (1 - \beta) i_v^*, \; \forall v \in V.$$

Let $z_v(t) = i_v(t) - i_v^*$ for all $v = 1, \cdots, n$. Therefore, we have

$$z_v(t+1) = -\beta z_v(t) + (1-\alpha) \prod_{(u,v) \in E} (1 - \gamma i_u^*) z_v(t) +$$

$$(1-\alpha) \left[ \prod_{(u,v) \in E} (1 - \gamma i_u^*) - \prod_{(u,v) \in E} (1 - \gamma i_u(t)) \right] (1 - i_v(t)).$$

Note that for each $v = 1, \cdots, n$,

$$\prod_{(u,v) \in E} (1 - \gamma i_u^*) - \prod_{(u,v) \in E} (1 - \gamma i_u(t))$$

$$= \gamma \sum_{(u',v) \in E} z_{u'}(t) \cdot \prod_{(u,v) \in E, \; u < u'} (1 - \gamma i_u^*) \cdot \prod_{(k,v) \in E, \; k > u'} (1 - \gamma i_k(t)),$$

we have

$$|z_v(t+1)| \leq h_v|z_v(t)| +$$
$$(1-\alpha)\gamma \sum_{(u,v)\in E} |z_u(t)| \left| \prod_{(u,v)\in E,\; u<u'} (1-\gamma i_u^*) \prod_{(k,v)\in E,\; k>u'} (1-\gamma i_k(t)) \right|$$
$$\leq h_v|z_v(t)| + (1-\alpha)\gamma \sum_{(u,v)\in E} |z_u(t)|.$$

Define the following comparison system

$$\theta_v(t+1) = h_v\theta_v(t) + (1-\alpha)\gamma \sum_{(u,v)\in E} \theta_u(t) \qquad (4.2)$$

with initial values $\theta(0)_v = |z_v(0)| \geq 0$ for $v = 1, \cdots, n$. This implies that $|z_v(t)| \leq \theta_v(t)$ holds for all $v = 1, \cdots, n$ and $t \geq 0$. Let $\Theta(t) = [\theta_1(t), \cdots, \theta_n(t)]^\top$. Then Eq. (4.2) has the following matrix form:

$$\Theta(t+1) = [H + \gamma(1-\alpha)\mathsf{A}]\,\Theta(t), \qquad (4.3)$$

where $\mathsf{A}$ is the adjacency matrix of the network graph. Since the graph is connected, the matrix $\mathsf{X} = H + \gamma(1-\alpha)\mathsf{A}$ is irreducible and nonnegative. The Perron-Frobenius theorem [Horn and Johnson 1985] says that the spectral radius of $\mathsf{X}$ is its largest eigenvalue in modulus. Therefore, condition (4.1) implies that all eigenvalues of the matrix $\mathsf{X}$ is less than 1 in modulus. This implies that system (4.2) converges to zero exponentially. Because $|z_v(t)| \leq \theta_v(t)$, $\lim_{t\to\infty}|z_v(t)| = 0$ holds for all $v = 1, \cdots, n$, i.e., $\lim_{t\to\infty} i_v(t) = i_v^*$ holds for all $v = 1, \cdots, n$ and the convergence is at an exponential pace. This completes the proof. □

The above theorem says that when the sufficient condition is satisfied, the spreading will converge to the equilibrium state (but not dying out) at an exponential pace, regardless of the number of initially infected nodes.

**Comparing the sufficient condition presented in [Chakrabarti et al. 2008] to ours.** Our result supersedes the sufficient condition for the case of $\alpha = 0$ presented in [Chakrabarti et al. 2008] because of the following corollary of Theorem 4.1:

COROLLARY 4.2. *In the case* $\alpha = 0$, *if*

$$\lambda_{1,\mathsf{A}} < \frac{\beta}{\gamma}, \qquad (4.4)$$

*the spreading dies out regardless of the number of initially infected nodes, i.e.,* $\lim_{t\to\infty} i_v(t) = 0$ *for all* $v = 1, \cdots, n$.

PROOF. Note that $\alpha = 0$ and $i_v^* = 0$ implies $H = (1-\beta)I_n$ in Theorem 4.1. So, condition (4.1) becomes $\lambda_{\max}[(1-\beta)I_n + \gamma A] < 1$, which is equivalent to $1 - \beta + \gamma\lambda_{\max}(A) < 1$, namely condition (4.4). Hence, the corollary holds. □

While Corollary 4.2 corresponds to the sufficient condition given in [Chakrabarti et al. 2008], meaning that our sufficient condition is strictly more powerful, the sufficient condition for the case of $\alpha = 0$ is not compatible with the more succinct sufficient condition for the case of $\alpha > 0$ (presented in Section 4.2). This also illustrates why the treatment in the case of $\alpha > 0$ is different.

**Caveat: The conditions are *sufficient* but not *necessary*.** [Chakrabarti et al. 2008] claimed through their Theorem 2 that $\tau = \frac{1}{\lambda_{1,\mathsf{A}}} > \frac{\beta}{\delta}$ (in their terminology), which is equivalent to the above condition (4.4), is also the *necessary* condition for the spreading to die out for the case of $\alpha = 0$. Now we point out that their proof is flawed and explain why. We use their terminology so that it is easier to check against their proof, which is presented in Appendix A of [Chakrabarti et al. 2008]. Their proof strategy is to prove "if the spreading dies out, then $\tau > \frac{\beta}{\delta}$." As shown in the derivation of their equation (A6), this requires to prove "if the system is asymptotically stable, then $\lambda_{i,\mathsf{S}} = 1 - \delta + \beta\lambda_{i,\mathsf{A}} < 1$ for $\forall i$, where $\mathsf{S} = \nabla g(\vec{0}) = \beta\mathsf{A} + (\vec{1} - \delta)\mathsf{I}$." This subsequently requires to prove "if the system is asymptotically stable at $\vec{P} = \vec{0}$, then the eigenvalues of $\nabla g(\vec{0})$ are less than 1 in absolute value." The proof of this claim was attributed to their Lemma 1, which however — as we now point out — cannot get through because the lemma actually states the opposite, namely "if the eigenvalues of $\nabla g(\vec{0})$ are less than 1 in absolute value, then the system is asymptotically stable at $\vec{P} = \vec{0}$." Therefore, the necessity proof in [Chakrabarti et al. 2008] is flawed.

Having pointed out the flaw, one would wonder whether it is possible to prove "if the system is asymptotically stable at $\vec{P} = \vec{0}$, then the eigenvalues of $\nabla g(\vec{0})$ are less than 1 in absolute value" or "$\tau > \frac{\beta}{\delta}$ in [Chakrabarti et al. 2008] is indeed the necessary condition for the spreading to die out." It turns out not to be the case. Conceptually, for a nonlinear dynamical system, when the largest eigenvalue of the Jacobin matrix at equilibrium equals to one exactly, it may not imply that the system is not asymptotically stable. For example, consider a simple system

$$x(t+1) = x(t) - \gamma(x(t))^2 \qquad (4.5)$$

with $0 < \gamma < 1/2$ and $x(0) \in [0, 1]$. Since $x(t)$ monotonically decreases and belongs to the interval $[0, 1]$ for all $t$, it must converge to a fixed point of the system. Since zero is the only fixed point of system (4.5), we conclude that $\lim_{t \to \infty} x(t) = 0$. Even though the modulus of the derivative of the right-hand side at zero is 1, the system converges to zero if the initial value is less than 1. In other words, their condition (as well as ours) is sufficient but not necessary.

Having pointed out that their proof strategy for necessity cannot get through, one may wonder whether the necessity can be proven with a different strategy. It turns out not to be the case, as we show here with a concrete counter-example that a sufficient condition is not necessary in general. Our counter-example below shows that (in our terminology) there exist graphs where the infection rate can still go to zero when $\lambda_{1,A} = \frac{\beta}{\gamma}$, which immediately violates the necessity of the sufficient condition. Let us consider a graph with only two nodes linked with a single edge. Then, we have the following dynamical system to describe the infection rate:

$$\begin{cases} i_1(t+1) = \gamma i_2(t)[1 - i_1(t)] + (1 - \beta)i_1(t) \\ i_2(t+1) = \gamma i_1(t)[1 - i_2(t)] + (1 - \beta)i_2(t). \end{cases} \qquad (4.6)$$

In this case, the adjacency matrix is $\begin{bmatrix} 0 & 1 \\ 1 & 0 \end{bmatrix}$, whose largest eigenvalue in modulus is $\lambda_{1,A} = 1$. So, under the condition $\lambda_{1,A} = \frac{\beta}{\gamma}$, we have $\beta = \gamma$. Suppose $0 < \gamma < 1/2$. Then, $0 \le i_2(t) \le 1$ implies $1 - \gamma - \gamma i_2(t) > 0$. Let $c \in [0, 1]$ such that $i_1(t) \le c$ and $i_2(t) \le c$ at time $t$, we have $\gamma - c\gamma \ge 0$ and thus

$$\begin{aligned} i_1(t+1) &= [1 - \gamma - \gamma i_2(t)]i_1(t) + \gamma i_2(t) \le [1 - \gamma - \gamma i_2(t)]c + \gamma i_2(t) \\ &= i_2(t)(\gamma - \gamma c) + c(1 - \gamma) \le c(\gamma - \gamma c) + c(1 - \gamma) = c - c^2\gamma. \end{aligned}$$

In a similar fashion we have $i_2(t+1) \le c - c^2\gamma$. Still considering the comparison system (4.5) with $x(0) = 1$, we observe that $i_1(t) \le x(t)$ and $i_2(t) \le x(t)$ hold. From the reasoning above, we have $\lim_{t \to \infty} x(t) = 0$. This implies that $\lim_{t \to \infty} i_1(t) = 0$ and $\lim_{t \to \infty} i_2(t) = 0$, which serves as a desired counter-example.

### 4.2. A More Succinct Epidemic Threshold

The above sufficient condition is general but not succinct. This motivates us to present the following less general, but more succinct, sufficient condition.

THEOREM 4.3. (a more succinct sufficient condition under which the spreading will become stable) *Let* $m = \max_{v \in V} \deg(v)$. *In the case* $\beta < (1-\alpha)(1 + (1-\gamma)^m)/2$, *if*

$$\lambda_{1,\mathsf{A}} < \frac{\alpha + \beta}{\gamma(1-\alpha)},$$

*then the spreading will become stable regardless of the initial number of infected nodes.*

*In the case* $\beta \geq (1-\alpha)(1 + (1-\gamma)^m)/2$, *if*

$$\lambda_{1,\mathsf{A}} < \frac{1 - \beta + (1-\alpha)(1-\gamma)^m}{\gamma(1-\alpha)},$$

*then the spreading will become stable regardless of the initial number of infected nodes.*

PROOF. Let $\psi = \max\left(|1 - \alpha - \beta|, \, |-\beta + (1-\alpha)(1-\gamma)^m|\right)$, and the other parameters be specified as above. Because

$$(1-\alpha)(1-\gamma)^m \leq (1-\alpha) \prod_{(u,v) \in E} (1 - \gamma i_u^*) \leq (1-\alpha),$$

we have

$$h_v \leq \max\{|1 - \alpha - \beta|, \, |-\beta + (1-\alpha)(1-\gamma)^m|\}.$$

So, the largest eigenvalue of $H + (1-\alpha)\gamma\mathsf{A}$ is smaller than that of $\psi I_n + (1-\alpha)\gamma\mathsf{A}$. Note that if

$$\psi + \gamma(1-\alpha)\lambda_{1,\mathsf{A}} < 1, \text{ or } \lambda_{1,\mathsf{A}} < \frac{1-\psi}{\gamma(1-\alpha)}, \tag{4.7}$$

then $\lambda_{\max}[H + (1-\alpha)\gamma\mathsf{A}] < 1$. This means, by applying Theorem 4.1, that the spreading will become stable regardless of the number of initially infected nodes.

Observe that in the case $\beta < (1-\alpha)(1 + (1-\gamma)^m)/2$, sufficient condition (4.7) becomes

$$\lambda_{1,\mathsf{A}} < \frac{\alpha + \beta}{\gamma(1-\alpha)}.$$

In the case $\beta \geq (1-\alpha)(1 + (1-\gamma)^m)/2$, sufficient condition (4.7) becomes

$$\lambda_{1,\mathsf{A}} < \frac{1 - \beta + (1-\alpha)(1-\gamma)^m}{\gamma(1-\alpha)}.$$

This completes the proof. □

**Comparing the sufficient condition given in [Li et al. 2007] to ours.** [Li et al. 2007] also presented a sufficient condition for the case $\alpha > 0$, namely

$$\lambda_{1,\mathsf{A}} < \frac{\alpha + \beta}{\gamma} \text{ and } \lambda_{n,\mathsf{A}} > \frac{\alpha + \beta - 2}{\gamma},$$

where $\lambda_{n,\mathsf{A}}$ is the smallest (in modulus) eigenvalue of $\mathsf{A}$. However, in the derivation of their sufficient condition they used the following approximation of Eq. (3.2)

$$i_v(t+1) \approx \alpha + (1 - \alpha - \beta)i_v(t) + \gamma \sum_{(u,v) \in E} i_u(t),$$

which omits all the nonlinear terms and is therefore quite coarse.

In contrast, our sufficient condition given in Theorem 4.3 is derived *without* using any approximation. As a result, our sufficient condition is advantageous because it is both *more concise* and *weaker*. Our sufficient condition is more concise because it says that $\lambda_{n,\mathsf{A}}$ is irrelevant; whereas, their sufficient condition unnecessarily involves $\lambda_{n,A}$. Our sufficient condition is weaker because when $\beta < (1-\alpha)(1+(1-\gamma)^m)/2$, $\lambda_{1,A} < \frac{\alpha+\beta}{\gamma}$ implies $\lambda_{1,A} < \frac{\alpha+\beta}{\gamma(1-\alpha)}$ but the latter does not imply the former.

### 4.3. Bounding the Node Infection Rate

The above theorems state sufficient conditions under which the spreading will become stable. Because we are unable to compute the equilibrium state for $\alpha > 0$, the next natural question is: Can we bound the nodes' infection rates? Now we address this question.

THEOREM 4.4. (Bounds of $i_v(t)$ regardless of the stability of the spreading) *Let* $\overline{\lim}_{t\to\infty} i_v(t)$ *denote the upper bound of the limit of* $i_v(t)$, *and* $\underline{\lim}_{t\to\infty} i_v(t)$ *denote the lower bound of the limit of* $i_v(t)$. *Then, we have*

$$\overline{\lim}_{t\to\infty} i_v(t) \le \theta_v^+ \text{ and } \underline{\lim}_{t\to\infty} i_v(t) \ge \theta_v^-,$$

*where*

$$\theta_v^+ = \frac{1-(1-\alpha)(1-\gamma)^{\deg(v)}}{\min\{1+\beta-(1-\alpha)(1-\gamma)^{\deg(v)},1\}},$$

$$\theta_v^- = \begin{cases} \frac{1-(1-\alpha)(1-\gamma\nu)^{\deg(v)}}{1+\beta-(1-\alpha)(1-\gamma\nu)^{\deg(v)}} & (1-\alpha)(1-\gamma\nu)^{\deg(v)} \ge \beta \\ ((1-\alpha)(1-\gamma\nu)^{\deg(v)}-\beta)\theta_v^+ + 1 - (1-\alpha)(1-\gamma\nu)^{\deg(v)} & otherwise \end{cases},$$

*with* $\nu = \min\{1-\beta, \alpha\}$.

PROOF. First, consider the upper bound. Because

$$i_v(t+1) \le [1-(1-\alpha)(1-\gamma)^{\deg(v)}](1-i_v(t)) + (1-\beta)i_v(t),$$

we have

$$i_v(t+1) - \theta_v^+ \le [(1-\alpha)(1-\gamma)^{\deg(v)} - \beta](i_v(t) - \theta_v^+).$$

In the case $(1-\alpha)(1-\gamma)^{\deg(v)} \ge \beta \ge 0$, there are two scenarios. If there exists some $t_0$ such that $i_v(t_0) < \theta_v^+$, then we have $i_v(t+1) \le \theta_v^+$ for all $t \ge t_0$. If $i_v(t) \ge \theta_v^+$ for all $t$, then

$$|i_v(t+1) - \theta_v^+| \le |[(1-\alpha)(1-\gamma)^{\deg(v)} - \beta]||i_v(t) - \theta_v^+|.$$

Because $|(1-\alpha)(1-\gamma)^{\deg(v)} - \beta| < 1$, we have

$$\lim_{t\to\infty}(i_v(t) - \theta_v^+) = 0.$$

So, we have $\overline{\lim}_{t\to\infty} i_v(t) \le \theta_v^+$ in both scenarios.

In the case $(1-\alpha)(1-\gamma)^{\deg(v)} < \beta$, we immediately have

$$\begin{aligned} i_v(t+1) &\le [(1-\alpha)(1-\gamma)^{\deg(v)} - \beta]i_v(t) + 1 - (1-\alpha)(1-\gamma)^{\deg(v)} \\ &\le 1 - (1-\alpha)(1-\gamma)^{\deg(v)}. \end{aligned}$$

Therefore, in both cases, we have $\overline{\lim}_{t\to\infty} i_v(t) \le \theta_v^+$.

Second, consider the lower bound. Since $1 - (1-\alpha)(1-\delta_v(t)) \ge \alpha$, we can conclude that $i_v(t+1) \ge \alpha(1-i_v(t)) + (1-\beta)i_v(t)$. The fact $i_v(t) \in [0,1]$ implies that the right-hand side above is greater than $\nu$. So, we have $i_v(t) \ge \nu$. Thus, we have

$$i_v(t+1) \ge [1-(1-\alpha)(1-\gamma\nu)^{\deg(v)}](1-i_v(t)) + (1-\beta)i_v(t).$$

In the case $(1-\alpha)(1-\gamma\nu)^{\deg(v)} \ge \beta$, we can rewrite the above inequality as

$$i_v(t+1) - \theta_v^- \ge [(1-\alpha)(1-\gamma\nu)^{\deg(v)} - \beta](i_v(t) - \theta_v^-).$$

We observe that there are two scenarios. If there exists some $t_1$ such that $i_v(t_1) > \theta_v^-$, then $i_v(t) \geq \theta_v^-$ holds for all $t \geq t_1$. If $i_v(t) \leq \theta_v^-$ holds for all $t \geq 0$, then

$$|i_v(t+1) - \theta_v^-| \leq |[(1-\alpha)(1-\gamma\nu)^{\deg(v)} - \beta]||i_v(t) - \theta_v^-|.$$

Because $|[(1-\alpha)(1-\gamma\nu)^{\deg(v)} - \beta]| < 1$, we have $\lim_{t\to\infty} |i_v(t+1) - \theta_v^-| = 0$. Therefore, we have $\underline{\lim}_{t\to\infty} i_v(t) \geq \theta_v^-$ in both scenarios.

In the case $(1-\alpha)(1-\gamma)^{\deg(v)} < \beta$, we immediately have

$$\underline{\lim}_{t\to\infty} i_v(t+1)$$
$$\geq \underline{\lim}_{t\to\infty}\left\{ [(1-\alpha)(1-\gamma\nu)^{\deg(v)} - \beta]i_v(t)) + 1 - (1-\alpha)(1-\gamma\nu)^{\deg(v)} \right\}$$
$$\geq [(1-\alpha)(1-\gamma\nu)^{\deg(v)} - \beta]\overline{\lim}_{t\to\infty} i_v(t)) + 1 - (1-\alpha)(1-\gamma\nu)^{\deg(v)}$$
$$\geq [(1-\alpha)(1-\gamma\nu)^{\deg(v)} - \beta]\theta_v^+ + 1 - (1-\alpha)(1-\gamma\nu)^{\deg(v)}.$$

Therefore, in both cases, we have $\underline{\lim}_{t\to\infty} i_v(t) \geq \theta_v^-$. This completes the proof. $\quad\Box$

The inequalities we use to prove Theorem 4.4 indicate that when the actual infection rate is significantly smaller than 1, the lower bound would be tighter; when the actual infection rate is close to 1, the upper bound would be tighter. This heuristics is confirmed by our simulation results. However, it is challenging to precisely pin down the infection rate threshold, above which the upper bound is tight and below which the lower bound is tight. We leave this to future research, while pointing out that a key difficulty comes from the fact hat the spreading may not be stable.

### 4.4. Applications of the Thresholds and Bounds

In addition to their theoretic significance, the thresholds and bounds have good applications, especially for *quantitatively* guiding the operation of tuning the parameters so as to achieve the desired outcomes.

First, the thresholds can be used to guide the adjustment of parameters so as to make the spreading stable. For this purpose, it is intuitive to decrease $\gamma$ (e.g., by imposing a more thorough examination on message packets) and/or $\alpha$ (e.g., by blocking access to potentially malicious websites), and/or increase $\beta$ (e.g., by deploying more powerful, and thus likely more expensive, defense tools). It would be less intuitive that the defender can also seek to decrease $\lambda_{1,A}$ by deleting edges and/or nodes. It would be far less intuitive that the optimal strategy is to delete edges/nodes so as to reduce $\lambda_{1,A}$ as much as possible, rather than to delete (for example) the largest-degree nodes as noted in [Chakrabarti et al. 2008]. In any case, how much adjustment is enough to assure that the spreading will become stable? The thresholds (more specifically, the inequalities) offered in Theorems 4.1 and 4.3 can be used to answer such questions quantitatively. Moreover, if the costs incurred due to the adjustment of parameters are known, then Theorems 4.1 and 4.3 also provide a basis for cost-effectively adjusting the parameters.

Second, consider the case of $\alpha = 0$. Suppose it is impossible or over-costly to make the spreading die out (e.g., because the parameters $\alpha$, $\beta$, $\gamma$, and $\lambda_{1,A}$ cannot be made arbitrarily small/large, while assuring that the network remains functioning). This means that the conditions given in Theorems 4.1 and 4.3 are not satisfied any more. Even if the spreading might not be stable, we can still utilize the bounds given in Theorem 4.4 to achieve some useful goals. On one hand, the upper bound can be used to estimate the worst-case infection rate and, perhaps more importantly, can be used to guide the tuning of parameters in the bound $\theta_v^+$ so as to reduce it as much as possible. Although it is not necessarily a tight estimation of the infection rate, it still would be useful because we now can control the worst-case infection rate under a desired level. On the other hand, the lower bound can be used to estimate in a sense the best-case scenario. Putting the bounds together, we further get a clue on how tight/loose the bounds are and, when the two bounds are close to each other, what is the small interval in which the infection rate resides. Finally, assuming the relevant cost functions are known, the tuning of parameters can also be made cost-effective. Such quantitative guidance would be always preferred to the qualitative intuitions.

Third, consider the case of $\alpha > 0$, which means that the spreading would never die out. In this case, even if we cannot make the spreading stable (similarly because it is either impossible or over-costly in practice as in the case of $\alpha = 0$), the above discussion on utilizing Theorem 4.4 to guide the adjustment of parameters would be still applicable here. Moreover, as shown in Section 5.1, the bounds hinted us to conduct some deeper analysis that leads to useful insights.

### 4.5. Confirming the Thresholds and Examining the Bounds via Simulation

We conducted a simulation study to confirm the analytical results by using two real-life complex network datasets obtained from `http://snap.stanford.edu/data/`.

— Epinions online social network dataset: This is a directed graph corresponding to a real-life online social network, where nodes represent people and arcs represent interactions between them. The graph has 75,879 nodes and 508,837 arcs with average node in- and out-degree 6.7059, maximal node in-degree 3,035, maximum node out-degree 1,801, and $\lambda_{1,A} = 106.53$.
— Enron email dataset: This is a graph representing Enron's internal email communications, where nodes represent employees and arcs represent email messages. Unlike the Epinions dataset, $(u, v) \in E$ always implies $(v, u) \in E$ in Enron dataset (i.e., the interactions between a pair of nodes, if any, are always mutual), which allows us to treat it as an undirected graph. The graph has 36,692 nodes and 367,662 edges with average degree 20.0404, maximal node degree 2,766, and $\lambda_{1,A} = 118.4177$.

Although not reported here, we note that both networks exhibit power-law degree distributions. The reason of using these datasets is that push- and pull-based epidemic spreading could take place in such networks.

Using the above datasets, we compare the simulation results (averaged over 500 independent runs) of the global mean infection rate $\bar{i}(t) = \frac{1}{|V|} \sum_{v \in V} i_v(t)$; the average (over all nodes) of the results obtained from Eq. (3.2); the average (over all nodes) of the bounds given in Theorem 4.4. The parameters are selected according to Theorem 4.3 so that the spreading will become stable. In all cases, we let 20% randomly picked nodes be initially infected. For the sake of better visual effect, we use higher resolutions for plotting figures whenever possible.

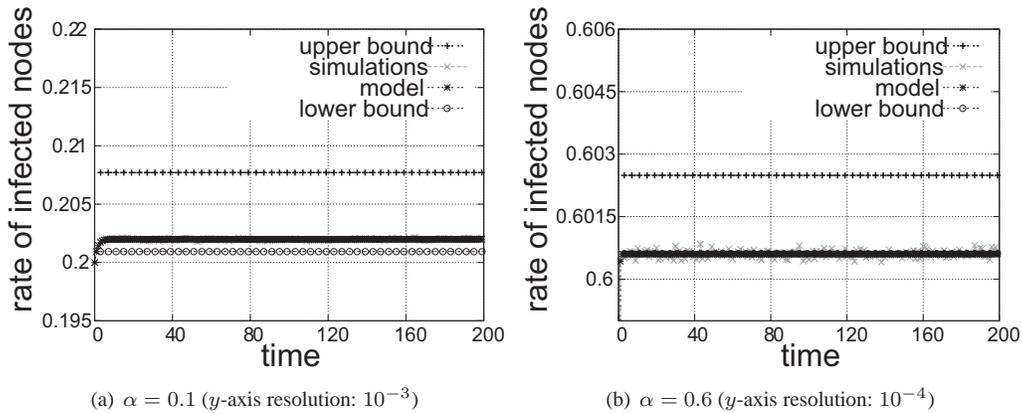

(a) $\alpha = 0.1$ ($y$-axis resolution: $10^{-3}$)  (b) $\alpha = 0.6$ ($y$-axis resolution: $10^{-4}$)

Fig. 2. The case of Epinions dataset ($\beta = 0.1$ and $\gamma = 0.004$)

Using the Epinions network dataset, Figure 2 plots the comparison of two parameter scenarios, in both cases the spreading will become stable. First, we note that the simulation result still oscillate a little bit in Figure 2(b) because the $y$-axis resolution is very high ($10^{-4}$). Indeed, we calculated

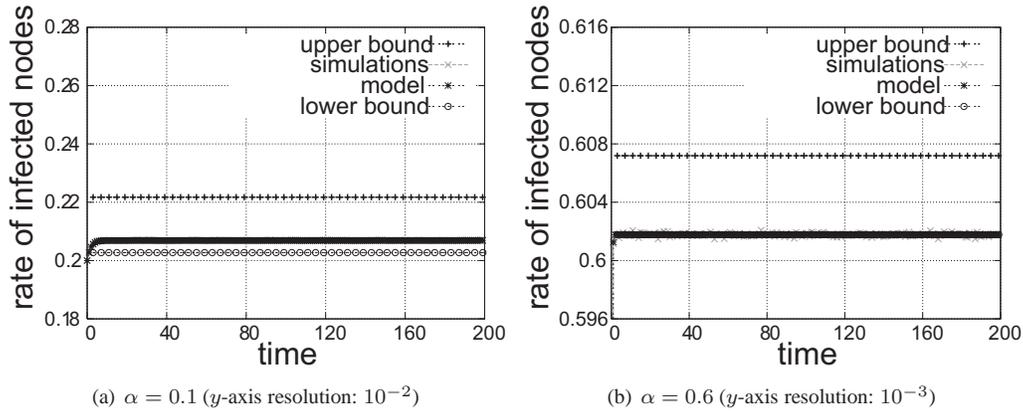

(a) $\alpha = 0.1$ ($y$-axis resolution: $10^{-2}$)    (b) $\alpha = 0.6$ ($y$-axis resolution: $10^{-3}$)

Fig. 3. The case of Enron email dataset ($\beta = 0.4$ and $\gamma = 0.001$)

the standard deviations of the simulation results in both scenarios, and found that the standard deviations are not visually noticeable (which explains why we did not plot the standard deviations). We interpret (at least partially) this phenomenon as follows. Since the initial conditions are random, the asymptotic limit of the infection rate, $i_v^*$, supposing it exists, is a random variable. Hence, the variance of the mean infection rate $\bar{i} = \frac{1}{n}\sum_{v=1}^{n} i_v^*$ is:

$$var(\bar{i}) = \frac{1}{n^2}var\left(\sum_{v=1}^{n} i_v^*\right) \leq \frac{2}{n^2}\sum_{v=1}^{n} var(i_v^*).$$

If $var(i_v^*)$ is bounded from above by some positive constant, then $var(\bar{i})$ converges to zero as $n \to \infty$, namely as the size of the network goes to infinity. Hence, the standard deviation, $\sqrt{var(\bar{i})}$, converges to zero. This would at least partially interpret why the observed standard deviations are very small. Second, even if in Figure 2(a) with $\alpha = 0.1$, we observe that the lower bounds are relatively tight. This in a sense confirms the aforementioned heuristics that when the infection rate is far from 1, the lower bound would be relatively tight. Third, it is noted that the simulation results and the model predictions are almost always lumped together.

Similar phenomena are observed in the case of Enron email dataset (Figure 3).

## 5. DEEPER INSIGHTS

In this section, we draw deeper insights related to the following three questions when the spreading is stable:

— Does $\deg(v)$ govern $i_v^* = \lim_{t\to\infty} i_v(t)$, namely what role do node-degrees play in terms of determining the nodes' infection rates (Section 5.1)?
— Under what conditions the mean field approach is reasonably accurate and thus applicable to arbitrary networks (Section 5.2)?
— When the mean field approach is reasonably accurate, how can we exploit it to infer the global mean infection rate through localized monitoring (Section 5.3)? This is especially important because it does not require knowledge of the parameters.

### 5.1. The Role of Node Degree on Node Infection Rate

Both the upper and lower bounds given in Theorem 4.4 are a function of $\deg(v)$. This suggests as a special case that $i_v^*$, the probability that node $v$ is infected when the spreading becomes stable, may be dependent upon $\deg(v)$. To characterize the impact of $\deg(v)$ on $i_v^*$, we start with an observation. Since we want to reveal the relation between $i_v^*$ and $\deg(v)$, we treat it as if $i_v^*$ depends only on $\deg(v)$. This allows us to consider $i^*(k)$, the probability any degree-$k$ node is infected after the spreading becomes stable. Suppose, for feasibility, all of $v$'s neighboring nodes have identical infection rate as $v$, namely that $i_u^* \approx i_v^*$ for all $u$ with $(u, v) \in E$. From master equation Eq. (3.2), we have

$$
\begin{aligned}
i_v^* &= \left[1 - (1 - \alpha) \prod_{(u,v) \in E} (1 - \gamma i_u^*)\right](1 - i_v^*) + (1 - \beta)i_v^* \\
&\approx \left[1 - (1 - \alpha)(1 - \gamma i_v^*)^k\right](1 - i_v^*) + (1 - \beta)i_v^*.
\end{aligned}
\tag{5.1}
$$

Hence, solution to the following equation

$$
\beta i^*(k) = \left[1 - (1 - \alpha)(1 - \gamma i^*(k))^k\right](1 - i^*(k))
\tag{5.2}
$$

is the probability that any degree-$k$ node is infected. In other words, solution to the following equation approximates $i^*(k)$:

$$
\beta x = [1 - (1 - \alpha)(1 - \gamma x)^k](1 - x)
\tag{5.3}
$$

for each $k = 1, 2, \cdots, \max_{v \in V}\{\deg(v)\}$. We observe that Eq. (5.3) can be numerically solved using (for example) Matlab.

In order to evaluate the above approximation-based analytic heuristics, we conducted a simulation study. In order to attest the applicability of the heuristics, we further used, in addition to the aforementioned Epinions and Enron datasets, the following datasets which vary in network topology and/or size:

— Oregon autonomous system dataset: This is an undirected graph of real-life network connections between Internet autonomous systems. The dataset is obtained from website `http://topology.eecs.umich.edu/data.html`. The graph has 11,461 nodes and 32,730 edges with average node degree 5.7115, maximal node degree 2,432, and largest eigenvalue $\lambda_{1,A} = 75.2407$. It also exhibits a power-law degree distribution.
— Power-law graphs: We used the Brite tool [Medina et al. 2001] to generate two synthetic undirected power-law graphs. The small one has 2,000 nodes and 5,997 edges with average node degree 5.997, maximal node degree 90, and $\lambda_{1,A} = 11.6514$. The large one has 37,000 nodes and 369,908 edges with average node degree 19.995, maximal node degree 1,530, and $\lambda_1 = 103.9362$.
— Random graphs: We used two synthetic undirected Erdos-Renyi graphs. The small one has 2,000 nodes and 6,001 edges with average node degree 6.001, maximal node degree 16, and $\lambda_{1,A} = 7.1424$. The large one has 37,000 nodes and 370,000 edges with average node degree 20, maximal node degree 40, and $\lambda_1 = 21.0894$.
— Regular graphs: The small synthetic undirected regular graph has 2,000 nodes and 6,000 edges with node degree 6 and $\lambda_{1,A} = 6$. The large synthetic undirected regular graph has 37,000 nodes and 370,000 edges with node degree 20 and $\lambda_1 = 20$.

In our simulations, we set $\alpha = 0.4$, $\beta = 0.6$ and $\gamma = 0.004$, which always satisfy the sufficient condition given in Theorem 4.3 and thus the spreading will become stable. We compare in Figures 4-5 the $i^*(k)$ obtained from simulation and the $i^*(k)$ obtained by solving Eq. (5.3). In each case, the $x$-axis represents degree $k = 1, \ldots, \max\{\deg(v)\}$, and the $y$-axis represents simulation-based $i^*(k)$ (+ in red color), which is obtained by averaging (over 100 simulation runs) the $i_v^*$'s of the degree-$k$ nodes after the spreading becomes stable, and the $i^*(k)$ obtained by numerically solving Eq. (5.3) ($\times$ in black color).

In the case of both real and synthetic power-law graphs (cf. Figure 4) the numerical results fit the simulation results pretty well, although not perfect. This means that the approximation-based Eq. (5.3) could be used in practice. On the other hand, we observe some intriguing phenomena.

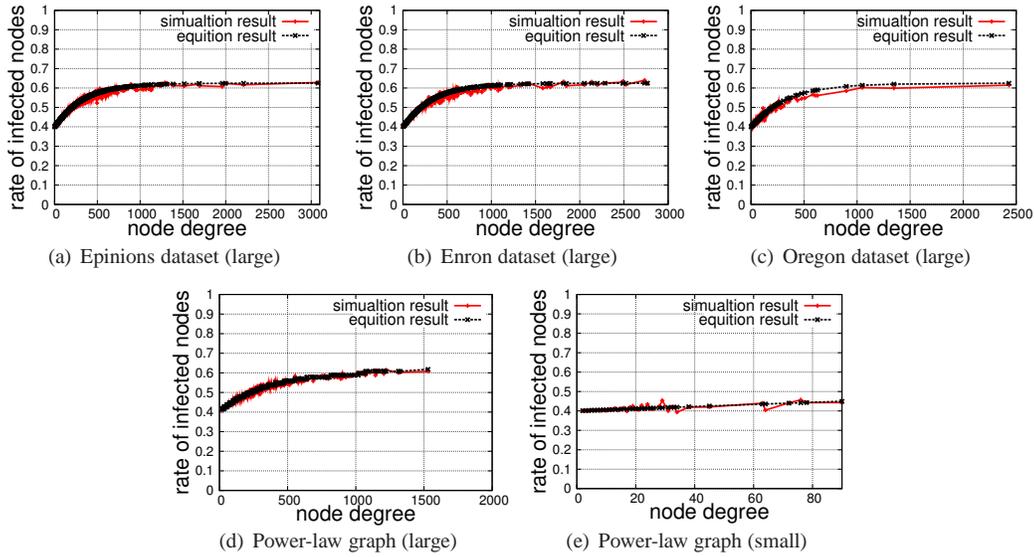

Fig. 4. On the accuracy of the approximation: Real and synthetic power-law graphs

— In the large real or synthetic power-law graphs (Figures 4(a)-4(d)), node degree $k$ has a non-linear factor on the nodes' infection rate $i^*(k)$. Whereas, for the small one (Figure 4(e)), node degree has a linear factor on $i^*(k)$. How can we interpret this discrepancy? We observe that the non-linearity appears to be proportional to the range of the node degrees. For example, the non-linearity in Figure 4(d) appears to be in between, so is its node degree interval. This phenomenon arguably can be explained as that a non-linear function could seem linear in a sufficiently small local region (e.g., degree range $[1, 250]$ in Figures 4(a)-4(d)).

— We observe that for sufficient large degree $k$ (e.g., when $k > 1,000$), $i^*(k)$ becomes almost flat. This can be explained as follows. We observe that the asymptotic limit $\lim_{k \to \infty} i^*(k) = 1/(1+\beta)$, which means that the variation of the slopes of the non-linear curves vanishes rapidly and thus the curves look flat because (for example) the limit value is 0.635 for $\beta = 0.6$.

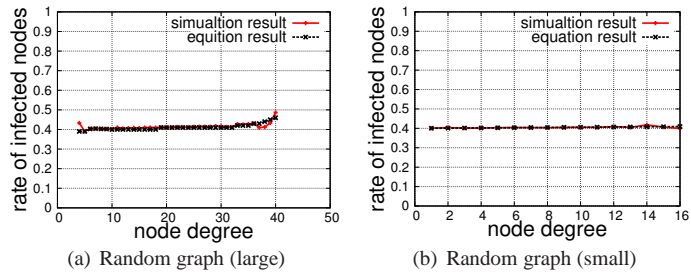

Fig. 5. On the accuracy of the approximation: Synthetic random graphs

In the case of synthetic Erdos-Renyi random graphs (Figure 5), the numerical results and the simulation results agree to a large extent. We observe that node degree $k$ has a linear factor on $i^*(k)$. Similar to the case of power-law graphs, this might be caused by the fact that the degrees are densely concentrated on a small interval, in which a non-linear function could be approximated as a linear one.

In the case of synthetic regular graphs, we report that the numerical results and the simulation results match nicely. We did not plot the figures because there is only a single point in each figure, which does not offer any extra value.

In summary, we draw the following:

INSIGHT 1. *Node degree plays an essential role in governing node infection rate. Moreover, Eq. (5.3) is a reasonably good approximation of $i^*(k)$, which means that one could use Eq. (5.3) to estimate $i^*(k)$ for given parameters.*

### 5.2. Condition under Which the Mean Field Approach Is Accurate

In this subsection, we will provide further analyses of the push- and pull-based epidemic models using the well-known mean field approach, which is a Statistical Physics method introduced for the purpose of studying stochastic particle systems [Chandler 1987]. Roughly speaking, the mean field approach is to use the mean (expectation) to stand for stochastic interactions. Mathematically speaking, it can be seen as the first-order approximation of random variables. To be concrete, we here give a simple example. Suppose that a variable $w$ is influenced by its random environment, which is described by a variable $x$, via a scheme $h$. That is, $w = h(x)$. By Taylor expansion, we have

$$w = h(x) = h(\langle x \rangle) + h'(\langle x \rangle)(x - \langle x \rangle) + o(|x - \langle x \rangle|),$$

where $\langle \cdot \rangle$ stands for the expectation (mean). By omitting the higher-order terms and conducting expectation on both sides, we have an approximation $\langle w \rangle = h(\langle x \rangle)$. This highlights the basic idea of the mean field approach: replacing a random field with its mean field.

The mean field approach also has become an important method for understanding complex dynamics. In general, this method is however not applicable in our setting because of the arbitrary network heterogeneity we accommodate. Therefore, it is natural to ask for conditions under which the method is applicable. In what follows we address this question.

We first derive a formula to describe $\bar{i}$, the global mean infection rate defined after the spreading becomes stable,

$$\bar{i} = \langle i_v^* \rangle = \frac{1}{|V|} \sum_{v \in V} i_v^* = \frac{1}{|V|} \sum_{v \in V} \lim_{t \to \infty} i_v(t)$$

in Erdos-Renyi random graph with edge probability $p$. Recall the equation for the fixed point of master equation Eq. (3.2):

$$\beta i_v^* = \left[ 1 - (1 - \alpha) \prod_{(u,v) \in E} (1 - \gamma i_u^*) \right] (1 - i_v^*).$$

So, we have

$$\beta \bar{i} = \left\langle \left[ 1 - (1 - \alpha) \prod_{(u,v) \in E} (1 - \gamma i_u^*) \right] (1 - i_v^*) \right\rangle.$$

We use the following approximation to derive the right-hand side term:

$$\left\langle \left[ 1 - (1 - \alpha) \prod_{(u,v) \in E} (1 - \gamma i_u^*) \right] (1 - i_v^*) \right\rangle \approx \left[ 1 - (1 - \alpha) \left\langle \prod_{(u,v) \in E} (1 - \gamma i_u^*) \right\rangle \right] (1 - \bar{i}).$$

In Erdos-Renyi random graphs with edge probability $p$, every pair of nodes are linked with probability $p$. So, we can approximate the following term as

$$\left\langle \prod_{(u,v)\in E} (1 - \gamma i_u^*) \right\rangle \approx (1 - \gamma p \bar{i})^n.$$

Let $\langle k \rangle$ represent the average degree. When $\gamma \langle k \rangle$ is sufficiently small, we have

$$\log[(1 - \gamma p \bar{i})^n] = n \log[1 - \gamma p \bar{i}] \approx -np\gamma\bar{i} = -\gamma \langle k \rangle \bar{i}.$$

Thus, we have

$$\left\langle \prod_{(u,v)\in E} (1 - \gamma i_u^*) \right\rangle \approx \exp(-\gamma \langle k \rangle \bar{i}).$$

So, we have

$$\beta \bar{i} \approx [1 - (1 - \alpha) \exp(-\langle k \rangle \gamma \bar{i})](1 - \bar{i}). \tag{5.4}$$

Its solution is $\bar{i}(\alpha, \beta, \gamma, \langle k \rangle)$. Although we are unable to derive its analytic expression, we can calculate the solution numerically.

The above discussion was made for Erdos-Renyi random graphs with edge probability $p$ with small $\gamma \langle k \rangle$. Can it be applied to other network topologies, perhaps under the condition $\gamma \langle k \rangle$ is small? To answer this question, we conduct simulation using some of the aforementioned datasets, which are selected with a certain balance on network size and topology (three large real datasets, three small synthetic datasets). Simulation results are based on the average of 50 runs.

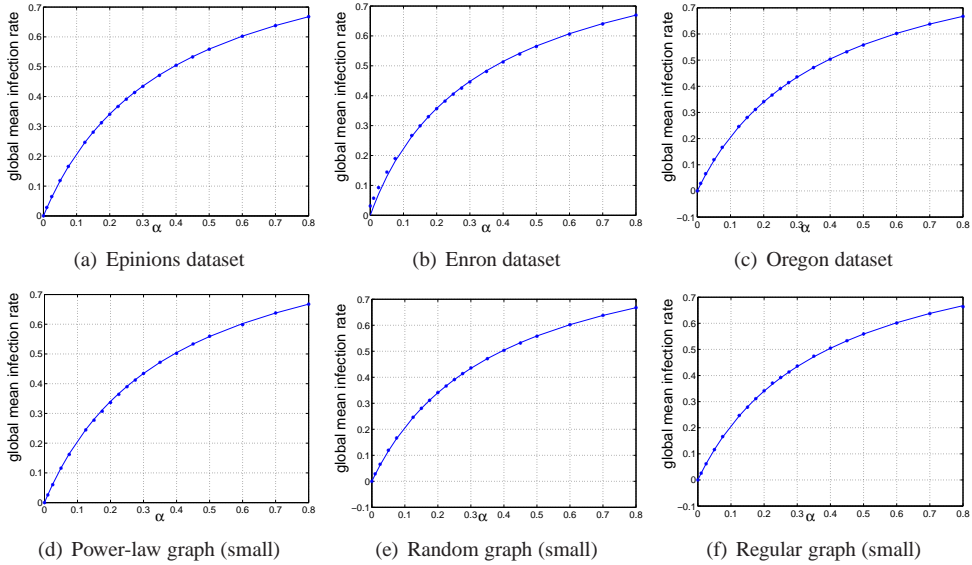

Fig. 6. Mean-field approximation via Eq. (5.4) (the curve) vs. simulation (the dots): $\beta = 0.4$ and $\gamma = 0.004$. Observe that $\bar{i}$ increases with respect to $\alpha$ in a concave fashion.

Figure 6 compares the $\bar{i}$ derived from the simulation results after the spreading becomes stable (the dots) and the numerical solution $\bar{i}$ of Eq. (5.4) with fixed $\beta = 0.4$ and $\gamma = 0.004$. In the case of the small synthetic Erdos-Renyi random graph, for which Eq. (5.4) was derived, we observe that the simulation result and the numerical result match perfectly. Moreover, for the other network graphs, they also match nicely. In all cases, the global mean infection rate $\bar{i}$ increases with respect to $\alpha$ in a concave fashion. This means that in order to significantly lower the global mean infection rate, $\alpha$ must be decreased substantially (e.g., from $\alpha = 0.6$ down to $\alpha = 0.2$). In other words, security against pull-based infection such as "drive-by download" attacks must be substantially enhanced; otherwise, the spreading will remain at a high level.

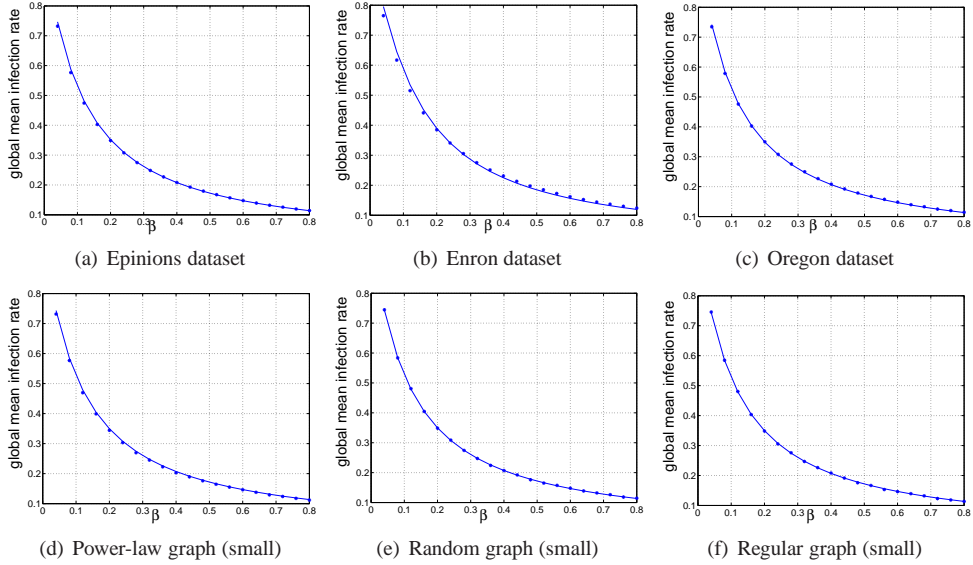

Fig. 7. Mean-field approximation via Eq. (5.4) (the curve) vs. simulation (the dots): $\alpha = 0.1$ and $\gamma = 0.004$. Observe that $\bar{i}$ decreases with respect to $\beta$ in a convex fashion.

Figure 7 compares the $\bar{i}$ derived from the simulation results after the spreading becomes stable (the dots) and the numerical solution $\bar{i}$ to Eq. (5.4) with fixed $\alpha = 0.1$ and $\gamma = 0.004$. In the case of Erdos-Renyi random graph, the simulation result and the numerical result match perfectly. For other network graphs, they also match well. In all cases, the global mean infection rate $\bar{i}$ decreases with respect to $\beta$ in a convex fashion. This means that in order to lower the global mean infection rate significantly, $\beta$ must be increased substantially (e.g., from $\beta = 0.1$ to $\beta = 0.5$). In other words, reactive security mechanisms against push- and pull-based infections must be substantially enhanced; otherwise, the spreading will remain at a relatively high level.

Figure 8 compares the $\bar{i}$ derived from the simulation results after the spreading becomes stable (the dots) and the numerical solution $\bar{i}$ to Eq. (5.4) with fixed $\alpha = 0.1$ and $\beta = 0.4$. In the case of random graph, the simulation result agrees with the numerical result. This also applies to the case of regular graph topology. In both cases, the global mean infection rate $\bar{i}$ increases with respect to $\gamma$ in a somewhat linear fashion. However, for power-law topologies, simulation result matches the numerical solution *only when* $\gamma$ (therefore, $\gamma \langle k \rangle$) is small; this confirms the condition under which the mean field approach is applicable to arbitrary networks. In any case, we observe that any enhancement in security mechanisms against spreading (e.g., blocking suspicious or malicious sessions) will always have an almost linear impact on the mean infection rate. This is quite different from the impact of $\alpha$ and $\beta$ because it states that network-based defense is always effective.

The above discussions lead us to draw the following:

INSIGHT 2. *The mean field approach is useful in the setting of arbitrarily heterogeneous networks when $\gamma \langle k \rangle$ is small. Under this circumstances, we found that the impacts of the parameters $\alpha$, $\beta$, and $\gamma$ are very different (as described above).*

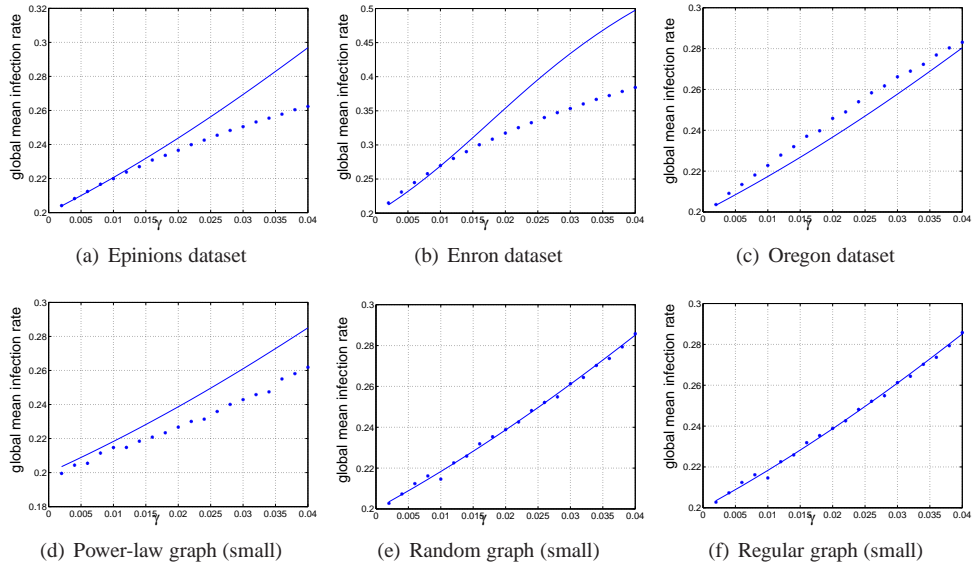

Fig. 8. Mean-field approximation via Eq. (5.4) (the curve) vs. simulation results (the dots): $\alpha = 0.1$ and $\beta = 0.4$. Observe that $\bar{i}$ increases with respect to $\gamma$ in a somewhat linear fashion.

### 5.3. Peeking into Global Infection via Localized Monitoring

**Estimating $\bar{i}$ via localized monitoring with unknown parameters**. In the above we showed that when $\gamma \langle k \rangle$ is small, solution to Eq. (5.4) is a good approximation of the global mean infection rate $\bar{i}$, Still, it requires to know the values of parameters $\alpha$, $\beta$ and $\gamma$. This naturally leads to another question: What can we say or (how) can we obtain/approximate $\bar{i}$ even if the parameters are not known? In theory, we can derive $\bar{i}$ by monitoring the *entire* network so as to count the number of infected nodes at every time $t$. This is clearly infeasible in practice. It then becomes natural to ask: How can we sample the network so as to approximate $\bar{i}$ with a reasonable precision? Because the states of the nodes are *not* independent of each other — rather they are correlated in a certain delicate fashion as governed by the master equation Eq. (3.2), a reasonable sampling strategy must take into consideration the evolution of the dynamics. In what follows we present such a result. To our surprise, the sample size can be constant and independent of the size of the network in question.

Recall that $i_v^*$ be the infection rate of node $v$ after the spreading becomes stable and $i^*(k)$ be the mean infection rate of the nodes with degree $k$. Then, we have

$$i^*(k) = \frac{1}{|\{v : \deg(v) = k\}|} \sum_{\deg(v)=k} i_v^* = \langle i_v^* | \deg(v) = k \rangle_v,$$

where $\langle \xi(u) | \Gamma \rangle_u$ denotes the conditional expectation of random variable $\xi(u)$ over $u \in V$ under condition $\Gamma$ (the subscript $u$ can be neglected if it is clear from the context). Recall Eq. (5.2), namely the approximation

$$\beta i^*(k) \approx [1 - (1 - \alpha)(1 - \gamma i^*(k))^k](1 - i^*(k)).$$

Thus, the global mean infection rate equals to the mean of $i^*(k)$ over the degree distribution:

$$\bar{i} = \langle i_v^* \rangle_v = \langle i^*(k) \rangle_k,$$

due to the property of conditional expectation that $\langle \xi \rangle = \langle \langle \xi | \eta \rangle \rangle$ for two arbitrary random variables $\xi$ and $\eta$. With linear approximation, we have

$$i^*(k) = i^*(\langle k \rangle) + \frac{di^*(k)}{dk}(k - \langle k \rangle) + o(k - \langle k \rangle),$$

which leads to

$$\bar{i} = \langle i^*(k) \rangle_k = i^*(\langle k \rangle) + \left\langle \frac{di^*(k)}{dk}(k - \langle k \rangle) \right\rangle_k + \langle o(k - \langle k \rangle) \rangle_k.$$

Neglecting the higher order terms, assuming

$$\left\langle \frac{di^*(k)}{dk}(k - \langle k \rangle) \right\rangle_k \approx \left\langle \frac{di^*(k)}{dk} \right\rangle_k \langle (k - \langle k \rangle) \rangle_k,$$

and because $\langle k - \langle k \rangle \rangle_k = 0$, we have the following approximation

$$\bar{i} \approx i^*(\langle k \rangle).$$

This leads us to draw the following:

INSIGHT 3. *We can measure (or approximate) $\bar{i}$, the global mean infection rate in a network, through the mean infection rate of the nodes with the average node degree of a network.*

To confirm the above insight, we use simulation to compare the global mean infection rate $\bar{i}$ and the mean infection rate of the average-degree nodes, denoted by $\bar{i}_{\text{avg}}$. We plot the results in Figure 9. The parameters used for simulation are $\alpha = 0.4$, $\beta = 0.4$ and $\gamma = 0.001$; in each of these cases, the spreading is stable.

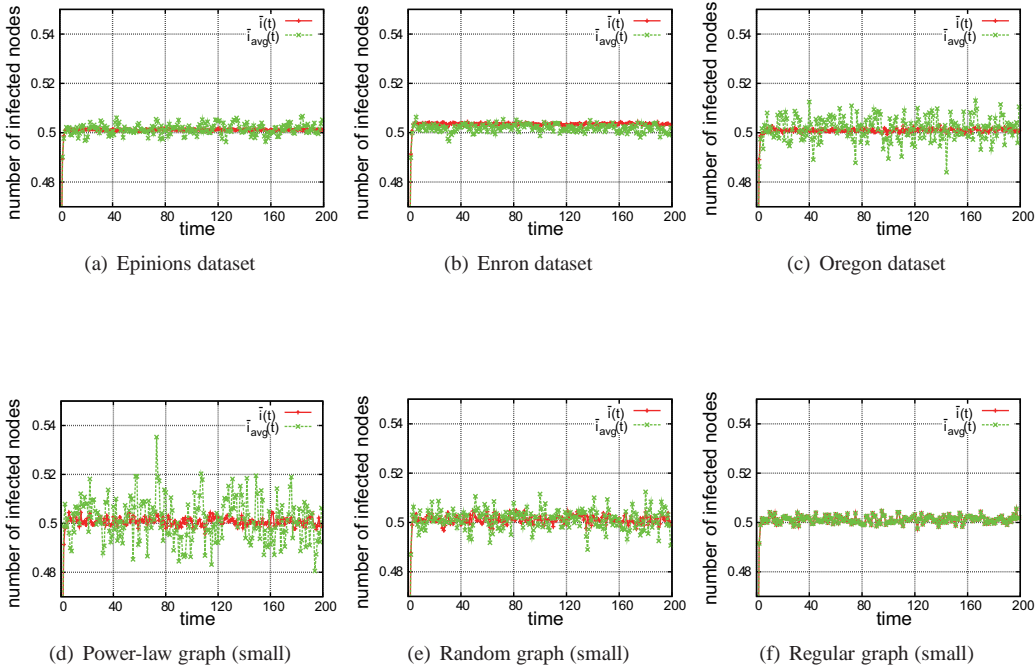

(a) Epinions dataset    (b) Enron dataset    (c) Oregon dataset

(d) Power-law graph (small)    (e) Random graph (small)    (f) Regular graph (small)

Fig. 9. $\bar{i}(t)$ vs. $\bar{i}_{\text{avg}}(t)$

Figure 9 does show that they are reasonably close to each other at the resolution of $10^{-2}$. The above insight is useful not only because the number of nodes with the average degree is much smaller than the total number of nodes, but also because we do not need to know the parameters $\alpha$, $\beta$ and $\gamma$. In other words, it states that monitoring the nodes with the average degree is sufficient to measure or approximate the global state of the network; this explains why we call it *localized monitoring*.

**Estimating $\bar{i}$ via localized monitoring with further reduced sample size**. While the above insight is already very useful, we ask a further deeper question: What if the number of nodes with the average degree is still large (e.g., thousands)? For example, in the case of Epinions dataset, there are 1,070 nodes (i.e., 1.41%) whose degrees equal to the average degree; in the case of Enron dataset, there are 2,212 nodes (i.e., 6.03%) whose degrees equal to the average degree. Monitoring and examining whether those nodes are compromised at every time $t$ is still costly. Is it possible to monitor even smaller number of nodes (ideally, constant number of nodes independent of the size of the network)? In what follows we answer this question affirmatively.

For the purpose of further reducing the number of nodes to monitor, we need to find further approximation to $i^*(\langle k \rangle)$. From the equations of the equilibrium:

$$\frac{\beta i_v^*}{1 - i_v^*} = 1 - (1 - \alpha) \prod_{(u,v) \in E} (1 - \gamma i_u^*), \; v \in V, \tag{5.5}$$

we know that the infection rate of node $v$ depends on those of its neighboring nodes. For the same reason, the infection rate of node $u$, which is a neighbor of node $v$, depends on the state of $u$'s neighboring nodes. Thus, we can regard $i_v^*$ as a function of $v$'s second-order neighborhood, which consists of nodes $r$ such that $r \neq v \wedge (r, u) \in E \wedge (u, v) \in E$. More precisely, we define the *second-order degree* of node $v$, denoted by $\mathsf{2deg}(v)$, as

$$\mathsf{2deg}(v) = \sum_{u:(u,v) \in E} \deg(u) - |\{w : (w, v) \in E \wedge (v, w) \in E\}|.$$

This means that when two different neighbors of node $v$ have a common neighbor $r$ besides $v$ itself, $r$'s contribution to the infection rate of $v$ should be counted twice. This leads us to consider the first and second order neighborhood of the nodes with the average degree $\langle k \rangle$. With the mean field approach, we have the following approximation:

$$\frac{\beta i_v^*}{1 - i_v^*} \approx 1 - (1 - \alpha)(1 - \gamma \langle i_u^* \rangle)^{\langle k \rangle}$$

and

$$\frac{\beta \langle i_u^* | (u, v) \in E \rangle}{1 - \langle i_u^* | (u, v) \in E \rangle} \approx \left\langle \frac{\beta i_u^*}{1 - i_u^*} \middle| (u, v) \in E \right\rangle$$

$$\approx 1 - (1 - \alpha)(1 - \gamma \langle i_r^* \rangle_{r:r \neq v \wedge (r,u) \in E \wedge (u,v) \in E})^{\langle \deg(u) | (u,v) \in E \rangle}.$$

As a result, we can further approximately regard $i_v^*$ with $\deg(v) = \langle k \rangle$ as a random variable over

$$\mathsf{2deg}(v) = [\langle \deg(u) | (u, v) \in E \rangle_u - 1] \deg(v).$$

Assuming $\langle i_r^* | (r, u) \in E, \; (u, v) \in E, \deg(v) = \langle k \rangle \rangle_v = i_v^*$ and letting $\tilde{i}_v = \langle i_u^* | (u, v) \in E \rangle$, we have

$$\frac{\beta i_v^*}{1 - i_v^*} = 1 - (1 - \alpha)(1 - \gamma \tilde{i}_v)^{\langle k \rangle} \quad \text{and} \quad \frac{\beta \tilde{i}_v}{1 - \tilde{i}_v} = 1 - (1 - \alpha)(1 - \gamma i_v^*)^{k'_v},$$

where

$$k'_v = \frac{\mathsf{2deg}(v)}{k}.$$

Given the mean degree $\langle k \rangle$, we can regard $i_v^*$ as a function of random variable $k_v'$, i.e., $i_v^* = f(k')$ for all $v$ with $\deg(v) = k$ and for some function $f(\cdot)$. By a discussion similar to the above, we have an approximation $\bar{i} \approx f(\langle k' \rangle) = f(\langle 2\deg(v) \rangle_{v:\deg(v)=k}/\langle \deg(v) \rangle)$. This means that we can select the nodes whose degrees equal to $\langle k \rangle$ and second-order degrees equal to $\langle 2\deg(v)|\deg(v) = \langle k \rangle \rangle$. This leads us to draw the following insight, which allows to further reduce the sample size.

INSIGHT 4. *We can approximate $\bar{i}$, the global mean infection rate, by only monitoring the nodes whose degrees equal to the average degree $\langle k \rangle$ and second-order degrees equal to $\langle 2\deg(v)|\deg(v) = \langle k \rangle \rangle$.*

The above insight statistically states that if there are no or few nodes whose degrees and second-order degrees satisfy the requirements, then we can select the nodes whose degrees and second-degree degrees are close to the respective average degrees. Because the insights do not specify how large the reduced sample should be, in what follows we use simulation to demonstrate how large the samples need to be.

Let 2avg-x denote the set of x nodes whose degrees equal to the average degree and second-order degrees are the closest to $\langle 2\deg(v)|\deg(v) = \langle k \rangle \rangle$. We note that there are two ways to approximate $\bar{i}$ via $\bar{i}_{\text{2avg-x}}$.

— Approximating $\bar{i}[t_0, t_1] = \frac{1}{t_1 - t_0 + 1}\sum_{t \in [t_0, t_1]} \bar{i}(t)$ as $\bar{i}_{\text{2avg-x}}[t_0, t_1] = \frac{1}{t_1 - t_0 + 1}\sum_{t \in [t_0, t_1]}\sum_{w \in \text{2avg-x}} i_w(t)$. This approximation is meaningful because the former captures the global mean infection rate averaged over time interval $[t_0, t_1]$ and the latter captures the mean infection rate of the x sampled nodes over the same period of time. Note that $t_0 > 0$ is used because we deal with the situation after the spreading becomes stable; our simulation result indicates that $t_0$ can be no greater than 10.

— Approximating $\bar{i}(t)$ via $\bar{i}_{\text{2avg-x}}[t_0, t] = \frac{1}{t - t_0 + 1}\sum_{t \in [t_0, t]}\sum_{w \in \text{2avg-x}} i_w(t)$. This approximation is useful and interesting because the latter captures the "history" of the small samples, which can actually approximate a certain index of the whole network.

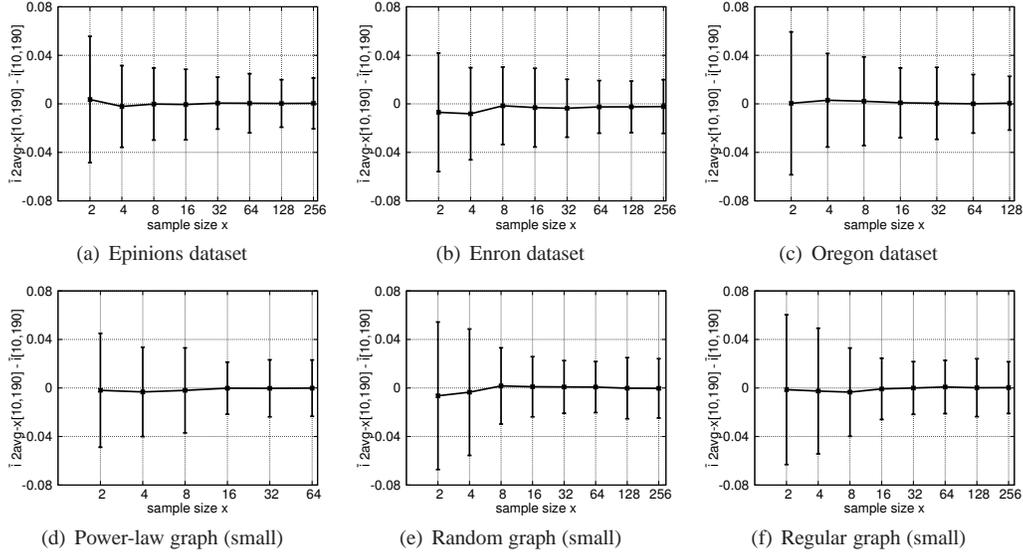

(a) Epinions dataset  (b) Enron dataset  (c) Oregon dataset

(d) Power-law graph (small)  (e) Random graph (small)  (f) Regular graph (small)

Fig. 10.  $\bar{i}_{\text{2avg-x}}[10, 190] - \bar{i}[10, 190]$

Corresponding to the first type of approximation mentioned above, Figure 10 plots $\bar{i}_{2\text{avg-x}}[t_0, t_1] - \bar{i}[t_0, t_1]$ over time interval $[t_0, t_1] = [10, 190]$ (with step-length 10) based on our simulation study, where $x \in \{2, 4, 8, 16, 32, 64, 128, 256\}$ nodes are selected as indicated above. We observe that in all the plotted network graphs, 16 nodes are enough to form a sufficiently accurate sample. What is surprising is that it seems a constant size of samples is sufficient, regardless of the size of the graph. It is an exciting future work to analytically explain this phenomenon.

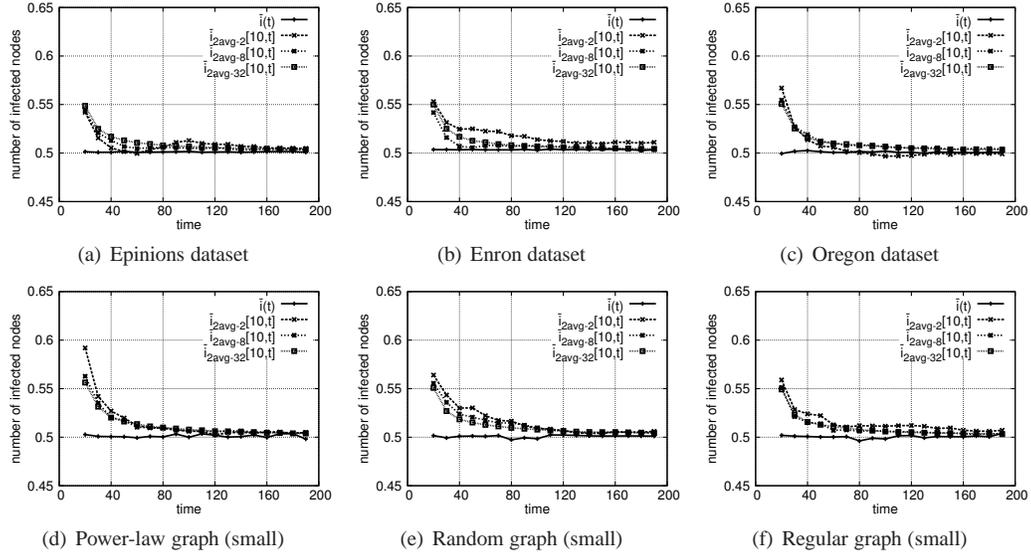

(a) Epinions dataset    (b) Enron dataset    (c) Oregon dataset

(d) Power-law graph (small)    (e) Random graph (small)    (f) Regular graph (small)

Fig. 11.    $\bar{i}_{2\text{avg-x}}[10, t]$ vs. $\bar{i}(t)$

Corresponding to the second type of approximation mentioned above, Figure 11 plots $\bar{i}_{2\text{avg-x}}[t_0, t]$ vs. $\bar{i}(t)$, where $t_0 = 10$, $t = 20, 30, \ldots, 200$, and $x \in \{2, 8, 32\}$ nodes are selected as indicated above. Note that the accuracy shown in Figure 11 is not as good as the accuracy shown in Figure 10. This is because Figure 10 corresponds to the average over time interval $[10, 190]$, where the comparison is between $\bar{i}[10, 190]$ and $\bar{i}_{2\text{avg-x}}[10, 190]$; whereas, Figure 11 corresponds to $\bar{i}(t)$ and $\bar{i}_{2\text{avg-x}}[10, t]$ for $t = 10, 20, \ldots, 190$. Nevertheless, as shown in Figure 11, for moderate $t$ (e.g., $t = 120$), the prediction of future is with reasonable accuracy. This leads to the following:

INSIGHT 5. *The accumulative observations obtained from monitoring certain constant number of nodes (selected as mentioned above; e.g., monitoring 32 nodes as shown in Figure 11) allow to estimate the current global mean infection rate $\bar{i}(t)$ with a pretty good accuracy.*

## 6. CONCLUSION AND FUTURE WORK

We rigorously investigated a push- and pull-based epidemic spreading model in arbitrary networks. We presented sufficient conditions or epidemic thresholds under which the spreading will become stable (but not dying out). The conditions supersede the relevant ones offered in the literature. Our investigation leads to further insights regarding the characterization of the role of node degree in governing node infection rate, the characterization of a condition under which the mean field approach is applicable in arbitrary networks, and a sampling strategy for selecting a small number of nodes to monitor so as to estimate the global mean infection rate *without* knowing the values of the parameters.

There are many exciting research problems from both theoretic and practical points of view. In addition to those mentioned in the text, here we highlight the following. First, how can we precisely pinning down the equilibrium states in the general case of $\alpha > 0$? Second, can we have tighter bounds especially when the spreading is not stable? Third, can we statistically prove that the sample size can be constant and thus independent of the size of the network (as indicated by our empirical study)? Fourth, can we design sampling algorithms to sample the nodes *without* knowing the network topology (our characterization demonstrates their existence but requires the knowledge of the global network)?

### Acknowledgement

We thank the anonymous reviewers for their comments that helped us improve the paper.

Shouhuai Xu and Li Xu were supported in part by grants sponsored by AFOSR, AFOSR MURI, ONR, and UTSA. Wenlian Lu was supported in part by the Foundation for the Author of National Excellent Doctoral Dissertation of PR China No. 200921, the National Natural Sciences Foundation of China under Grant No. 60804044, and the Shanghai Pujiang Program No. 08PJ14019.

### REFERENCES

ANDERSON, R. AND MAY, R. 1991. *Infectious Diseases of Humans*. Oxford University Press.

BAILEY, N. 1975. *The Mathematical Theory of Infectious Diseases and Its Applications*. 2nd Edition. Griffin, London.

BARABASI, A. AND ALBERT, R. 1999. Emergence of scaling in random networks. *Science 286*, 509–512.

BARRAT, A., BARTHLEMY, M., AND VESPIGNANI, A. 2008. *Dynamical Processes on Complex Networks*. Cambridge University Press.

CALLAWAY, D. S., NEWMAN, M. E. J., STROGATZ, S. H., AND WATTS, D. J. 2000. Network robustness and fragility: Percolation on random graphs. *Phys. Rev. Lett. 85*, 25, 5468–5471.

CHAKRABARTI, D., WANG, Y., WANG, C., LESKOVEC, J., AND FALOUTSOS, C. 2008. Epidemic thresholds in real networks. *ACM Trans. Inf. Syst. Secur. 10*, 4, 1–26.

CHANDLER, D. 1987. *Introduction to modern statistical mechanics*. Oxford University Press.

CHEN, Z. AND JI, C. 2005. A self-learning worm using importance scanning. In *Proceedings of the 2005 ACM Workshop on Rapid Malcode (WORM'05)*. 22–29.

DEMERS, A., GREENE, D., HAUSER, C., IRISH, W., LARSON, J., SHENKER, S., STURGIS, H., SWINEHART, D., AND TERRY, D. 1987. Epidemic algorithms for replicated database maintenance. In *PODC*. 1–12.

FALOUTSOS, M., FALOUTSOS, P., AND FALOUTSOS, C. 1999. On power-law relationships of the internet topology. In *Proc. of SIGCOMM'99*. 251–262.

GANESH, A., MASSOULIE, L., AND TOWSLEY, D. 2005. The effect of network topology on the spread of epidemics. In *Proceedings of IEEE Infocom 2005*.

HETHCOTE, H. 2000. The mathematics of infectious diseases. *SIAM Rev. 42*, 4, 599–653.

HORN, R. AND JOHNSON, C. 1985. *Matrix Analysis*. Cambridge University Press.

KARP, R., SCHINDELHAUER, C., SHENKER, S., AND VÖCKING, B. 2000. Randomized rumor spreading. In *FOCS*. 565–574.

KEMPE, D. AND KLEINBERG, J. 2002. Protocols and impossibility results for gossip-based communication mechanisms. 471–480.

KEMPE, D., KLEINBERG, J., AND DEMERS, A. 2001. Spatial gossip and resource location protocols. In *STOC*. 163–172.

KEPHART, J. AND WHITE, S. 1991. Directed-graph epidemiological models of computer viruses. In *IEEE Symposium on Security and Privacy*. 343–361.

KEPHART, J. AND WHITE, S. 1993. Measuring and modeling computer virus prevalence. In *IEEE Symposium on Security and Privacy*. 2–15.

KERMACK, W. AND MCKENDRICK, A. 1927. A contribution to the mathematical theory of epidemics. *Proc. of Roy. Soc. Lond. A 115*, 700–721.

LI, X., PARKER, P., AND XU, S. 2007. Towards quantifying the (in)security of networked systems. In *21st IEEE International Conference on Advanced Information Networking and Applications (AINA'07)*. 420–427.

MCKENDRICK, A. 1926. Applications of mathematics to medical problems. *Proc. of Edin. Math. Soceity 14*, 98–130.

MEDINA, A., LAKHINA, A., MATTA, I., AND BYERS, J. 2001. Brite: An approach to universal topology generation. In *Proc. International Symposium in Modeling, Analysis and Simulation of Computer and Telecommunication Systems (MASCOTS'01)*. 346–356.

MOLLOY, M. AND REED, B. 1995. A critical point for random graphs with a given degree sequence. *Random Struct. Algorithms 6*, 161–179.

MOLLOY, M. AND REED, B. 1998. The size of the giant component of a random graph with a given degree sequence. *Comb. Probab. Comput. 7*, 295–305.

MORENO, Y., PASTOR-SATORRAS, R., AND VESPIGNANI, A. 2002. Epidemic outbreaks in complex heterogeneous networks. *European Physical Journal B 26*, 521–529.

NEWMAN, M. 2003. The structure and function of complex networks. *SIAM Review 45*, 167.

NEWMAN, M. 2007. Component sizes in networks with arbitrary degree distributions. *Phys. Rev. E 76*, 4, 045101.

NEWMAN, M. 2010. *Networks: An Introduction*. Oxford University Press.

NEWMAN, M. E. J., STROGATZ, S. H., AND WATTS, D. J. 2001. Random graphs with arbitrary degree distributions and their applications. *Phys. Rev. E 64*, 2, 026118.

PASTOR-SATORRAS, R. AND VESPIGNANI, A. 2001. Epidemic dynamics and endemic states in complex networks. *Physical Review E 63*, 066117.

PASTOR-SATORRAS, R. AND VESPIGNANI, A. 2002. Epidemic dynamics in finite size scale-free networks. *Physical Review E 65*, 035108.

PROVOS, N., MCNAMEE, D., MAVROMMATIS, P., WANG, K., AND MODADUGU, N. 2007. The ghost in the browser analysis of web-based malware. In *Proceedings of the First Workshop on Hot Topics in Understanding Botnets (HotBots'07)*.

SHAH, D. 2009. Gossip algorithms. *Foundations and Trends in Networking 3*, 1, 1–125.

WANG, Y., CHAKRABARTI, D., WANG, C., AND FALOUTSOS, C. 2003. Epidemic spreading in real networks: An eigenvalue viewpoint. In *Proc. of the 22nd IEEE Symposium on Reliable Distributed Systems (SRDS'03)*. 25–34.

WILF, H. 1994. *generatingfunctionology*. Academic Press, Inc.

WILLINGER, W., ALDERSON, D., AND DOYLE, J. 2009. Mathematics and the internet: A source of enormous confusion and great potential. *Notices of the American Mathematical Society 56*, 5, 286–299.

YOSHIDA, K. 1971. *Functional Analysis and Its Applications*. Springer.

ZOU, C., GAO, L., GONG, W., AND TOWSLEY, D. 2003. Monitoring and early warning for internet worms. In *Proc. of ACM-CCS 03*, V. Atluri, Ed. ACM, ACM Press, Washington D.C., USA, 190–199.

ZOU, C., GONG, W., AND TOWSLEY, D. 2002. Code red worm propagation modeling and analysis. In *Proc. of the 9th ACM Conference on Computer and Communications Security*. 138–147.